\documentclass[jcp,twocolumn,showpacs,superscriptaddress,groupedaddress]{revtex4}  
\usepackage{graphicx}  
\usepackage{dcolumn}   
\usepackage{bm}        
\usepackage{amssymb}   
\usepackage{color}
\usepackage{mathrsfs}  

\begin{document}

\title{Reduced strength and extent of dynamic heterogeneity in a strong glass former as compared to fragile glass formers}
\author{Hannah Staley}
\affiliation{Department of Physics, Colorado State University, Fort Collins, Colorado 80523, USA}
\author{Elijah Flenner}
\author{Grzegorz Szamel}
\affiliation{Department of Chemistry, Colorado State University, Fort Collins, Colorado 80523, USA}

\date{\today}

\begin{abstract}
We examined dynamic heterogeneity in a model tetrahedral network glass-forming liquid. We used four-point correlation functions to extract dynamic correlation lengths $\xi_4^a(t)$ and susceptibilities $\chi_4^a(t)$ corresponding to structural relaxation on two length scales $a$. One length scale corresponds to structural relaxation at nearest neighbor distances and the other corresponds to relaxation of the tetrahedral structure. We find that the dynamic correlation length $\xi_4^{a}$ grows much slower with increasing relaxation time than for model fragile glass formers. We also find that $\chi_4^a \sim (\xi_4^a)^z$ for a range
of temperatures, but $z < 3$ at the lowest temperatures examined in this study. 
However, we do find evidence that the temperature where Stokes-Einstein violation begins marks a temperature where there is a change in the character of dynamically heterogeneous regions. Throughout the paper, we contrast the structure and dynamics of a strong glass former with that of a representative fragile glass former. 
\end{abstract}

\maketitle

\section{Introduction}

The dramatic increase of the structural relaxation time, $\tau_\alpha$, upon a small change in temperature without any major structural change is a defining characteristic of glassy dynamics. One can describe the temperature dependence of $\tau_\alpha$ using $\tau_\alpha = \tau_0 \exp[E(T)/T]$, where $E(T)$ is a possibly temperature dependent activation energy. Angell \cite{Angell} classified glass forming liquids according to the temperature dependence of their activation energy. Liquids with a temperature-independent activation energy, $E(T) = E_0$, were termed strong and those with an activation energy $E(T)$ increasing with decreasing temperature were termed fragile. The Arrhenius temperature dependence of a strong glass former implies that there is a single energy barrier for relaxation and a growing $E(T)$ implies a growing energy barrier. A natural mechanism for this growing energy barrier is an increase in the number of particles that have to move cooperatively to facilitate structural relaxation as the relaxation time increases, and much research has been devoted to the search of a length scale that can be associated with these cooperatively rearranging regions. 

Spatially correlated heterogeneous dynamics, dynamic heterogeneity, emerged as a candidate for the length scale associated with an increasing activation energy \cite{Ediger,BB}. It was found that there are clusters of particles that move much slower and much faster than expected from a Gaussian distribution of displacements, and the size of these clusters increases with increasing relaxation time. To classify the size and shape of these dynamically heterogeneous clusters, one approach is to calculate a four-point structure factor $S_4(q;t)$ \cite{LSSG} that involves a weight function $w_n(t)$ associated with mobility of particle $n$ over a period of time equal to $t$. Common choices for the weight function are the real part of the microscopic self-intermediate scattering function \cite{Berthier2004,FS2015} and an overlap function that is defined to be zero if a particle moves beyond a specified fraction of the particle diameter after a time $t$, \cite{LSSG}. The small $q$ behavior of $S_4(q;t)$ can be analyzed to determine a characteristic length of dynamic heterogeneity $\xi_4(t)$ \cite{LSSG,FS2010}. The susceptibility, $\chi_4(t) = \lim_{q \rightarrow 0} S_4(q;t)$, is related to the number of particles whose mobility is correlated. Calculation of $S_4(q;t)$ is straightforward in a simulation but it is impossible in most experiments where individual particles cannot be tracked (and it is difficult even if the particles can be tracked). However, Berthier \textit{et al.} \cite{BSci} demonstrated that an approximation for $\chi_4(t)$ can be obtained from experiments by considering dynamic fluctuations in different statistical ensembles. 

Berthier \textit{et al.}'s observation was that $\chi_4(t)$ consists of two parts, $\chi_4(t) = \chi_4(t)|_{NVE} + \mathcal{X}(t)$, where $\chi_4(t)|_{NVE}$ represents the fluctuations of $w_n(t)$ calculated in the micro-canonical ensemble (with constant number of particles and constant energy) and $\mathcal{X}(t)$ represents the correction term accounting for the fluctuations absent in the micro-canonical ensemble. One correction term is derived by considering the transformation from the micro-canonical to the canonical ensemble, and this term is given by $\chi_{4,T}(t) = k_B T^2 \chi_T(t)^2/c_v$ where $c_v$ is the specific heat at constant volume, $\chi_T(t) = \partial \left< w(t)\right>/\partial T$, and $\left<w(t)\right>$ is the average of the weight function $w_n(t)$ \cite{BBBKMR1,BBBKMR2}. If one considers the real part of the self-intermediate scattering function as the weight function, and assumes that time-temperature superposition at least approximately holds, then the correction term is approximately $k_B E_0^2/(T^2 c_v)$ for a strong glass former.  Therefore, in the low temperature limit the susceptibility diverges, which implies a diverging number of particles whose mobility is correlated, and thus a diverging correlation length. However, it is clear that for a fragile liquid the growth of the susceptibility $\chi_{4,T}(t)$ would be different than for a strong glass former.  

 We recently examined dynamic heterogeneity in five different fragile liquids \cite{FSS}, and found some universal behavior in all the systems studied. We found several scaling relationships if we rescaled  relaxation times by the relaxation time of the liquid corresponding to the temperature, $T_s$, at which  the Stokes-Einstein relation is violated. We also found that $T_s$ corresponds to the temperature where dynamically correlated regions become compact, and Hocky \textit{et al.}\ \cite{HBKR} found evidence for a change in shape of dynamically correlated regions around $T_s$. We note that a connection  between Stokes-Einstein violation and the shapes of dynamically correlated regions has been recently made in an two-dimensional colloidal mixture with a wall of pinned particles \cite{NGSG} and a quasi-two-dimensional colloidal system of ellipsoids \cite{MG}.  Therefore, Stokes-Einstein violation appears to be related to the shape and size of dynamically heterogeneous regions. 
 
In this work we examine dynamic heterogeneity for a model strong glass former, and focus on what is different between a strong glass former and the features we found for several fragile glass formers. We describe the model and simulations in Sec.~\ref{sec:sim}. After we examine the structure, average dynamics, and Stokes-Einstein violation in Sec.~\ref{sec:dyn}, we examine dynamic heterogeneity in Sec.~\ref{sec:dynhet}. We find that the strength and size of the dynamic heterogeneity is very different in a strong glass than in fragile glass formers. In Sec.~\ref{sec:sum} we summarize our results and draw some conclusions.

\section{Simulations}
\label{sec:sim}

A simple and convenient model for a strong glass-forming liquid was developed by Coslovich and Pastore \cite{ntw}. We will refer to this model as the CP model. The model was designed to model a tetrahedral network glass former, SiO$_2$, and thus Coslovich and Pastore compared it to another model for SiO$_2$, the BKS model \cite{BKS}, finding reasonable agreement. 
The model of Coslovich and Pastore does not include long range electrostatic forces, and thus it is much less computationally expensive than other models of strong liquids. 

We simulated the CP model strong glass-forming liquid. It is a 2:1 binary mixture of particles interacting with the following potential,
\begin{equation}
\label{eq:V}
U_{\alpha\beta}(r) = \epsilon_{\alpha\beta} \left[\left(\frac{\sigma_{\alpha\beta}}{r}\right)^{12} - (1 - \delta_{\alpha\beta})\left(\frac{\sigma_{\alpha\beta}}{r}\right)^6\right], 
\end{equation}
where $\delta_{\alpha\beta}$ is the Kronecker delta, and $\alpha, \beta = 1,2$. We use the reduced units of $\sigma_{11}$, $\epsilon_{11}$, and $\sqrt{m_1 \sigma_{11}^2/\epsilon_{11}}$ for length, energy, and time, respectively. The parameters of $U_{\alpha \beta}(r)$ are given by $\epsilon_{12} = 24\epsilon_{11}$, $\epsilon_{22} = \epsilon_{11}$, $\sigma_{12} = 0.49\sigma_{11}$, $\sigma_{22} = 0.85\sigma_{11}$, and The mass ratio is $m_2/m_1 = 0.57/1.0$. Following the work of Coslovich and Pastore \cite{ntw}, a smoothing function is appended to the potential at $r = 2.2\sigma_{\alpha\beta}$ \cite{smooth}.  We performed simulations of $N_1 = 9000$ and $N_2 = 18000$ at a particle density of $\rho = N/V = 1.655$. We used a time step of $\delta t = 0.001$ for $T \ge 0.5$ and a time step of $\delta t = 0.004$ for $T < 0.5$.

We ran our simulations using LAMMPS \cite{L_o,L_a}, (Large-scale Atomic/Molecular Massively Parallel Simulator). We equilibrated at all temperatures for at least $100\tau_{\alpha}^{0.35}$ ($\tau_{\alpha}^{0.35}$ is a relaxation time, which will be defined precisely in Sec. \ref{sec:dyn}). We then ran 4 independent NVE runs at each temperature for at least $100\tau_{\alpha}^{0.35}$. Many of the longer runs were run on the ISTeC Cray Model XE6 at Colorado State University.
 
We compare some results to a model fragile glass former of repulsive harmonic spheres (HARM), where the potential is given by 
\begin{equation}
\label{eq:V_harm}
U_{\alpha\beta}(r) = \frac{\epsilon}{2} \left( 1 - \frac{r}{\sigma_{\alpha\beta}} \right)^2,
\end{equation}
for $r < \sigma_{\alpha\beta}$ and $U_{\alpha\beta}(r) = 0$ for $r \ge \sigma_{\alpha\beta}$. The results for the harmonic sphere system are presented  using the reduced units of $\sigma_{11}$, $10^{-4} \epsilon$, $\sqrt{\epsilon \sigma_{11}^2/m}$ for length, temperature, and time, respectively. The harmonic sphere system is a 50:50 binary mixture with equal masses $m$. The potential parameters are $\sigma_{22} = 1.4\sigma_{11}$ and $\sigma_{12} = 1.2\sigma_{11}$. The number density is $\rho = 0.675$. We simulated a systems of $N = 10,000$, 40,000, and 100,000 particles. See Refs.~\cite{FS13,FS15} for more details. 

\section{Structure and Dynamics}
\label{sec:dyn}

We begin our study by looking at the structure and average dynamics of the CP model, and make some comparisons with the harmonic sphere fragile glass former. As is common in network forming liquids, in the structure factor there is a peak at low $q$ for the CP model that has no analog for the fragile glass former. This additional peak motivates us to examine the dynamics at two length scales, one associated with the nearest neighbor distance and one associated with the peak at smaller $q$. We then examine the average dynamics and demonstrate that an Arrhenius law does fit the relaxation time and the diffusion coefficients well. We finish by examining Stokes-Einstein violation in this system and obtain the temperature where the Stokes-Einstein relaxation is violated. 

\begin{figure}
\includegraphics[scale=0.6]{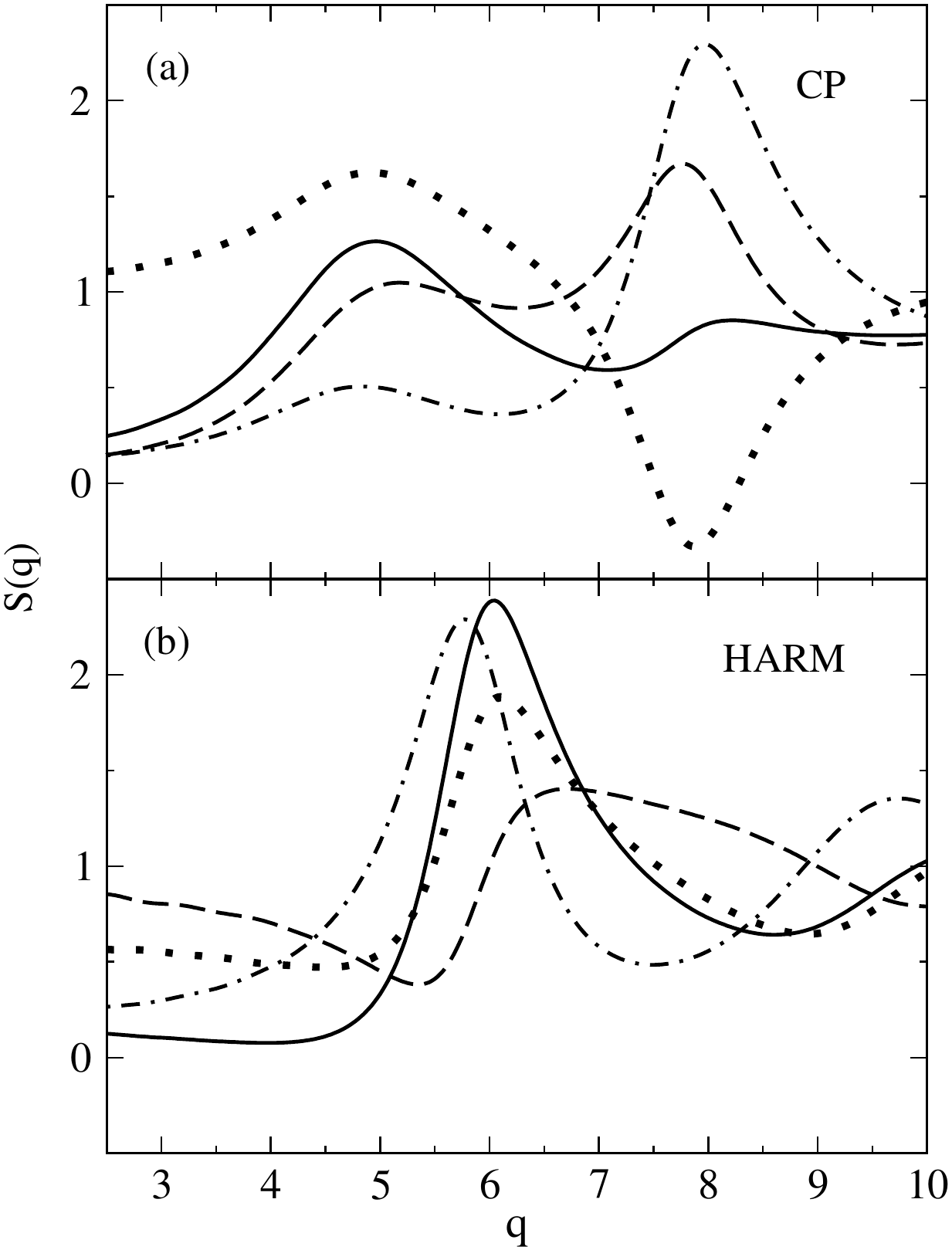}
\caption{\label{fig:Sq}The structure factor for the CP model at $T=0.42$ (a) and for the harmonic sphere system at $T=10$ (b). The solid lines are the total structure factors, the dashed lines are the partial structure factors $S_{11}(q)$, the dotted lines are the partial structure factors $S_{12}(q)$, and the dash-dotted lines are the partial structure factors $S_{22}(q)$. The peaks around $q=5$ for the CP model are due to intermediate range tetrahedral order.}
\end{figure}

Shown in Fig.~\ref{fig:Sq}a are the partial structure factors
\begin{equation}
\label{eq:Sqreal}
S_{\alpha\beta} (q) = \frac{1}{\sqrt{N_{\alpha}N_{\beta}}} \left< \sum_{m=1}^{N_{\alpha}} \sum_{n=1}^{N_{\beta}} e^{i\mathbf{q}\cdot (\mathbf{r}_n - \mathbf{r}_m)} \right>,
\end{equation}
where $\mathbf{r}_n$ is the position of particle $n$, and the total structure factor $S(q) = N^{-1}[N_1 S_{11}(q) + N_2 S_{22}(q) + 2 \sqrt{N_1 N_2} S_{12}(q)]$ for the CP model. There are peaks at $q = 5.0$ and 8.2 for the total structure factor (solid line). The peak at $q=5.0$ is a result of intermediate range order that is typical in tetrahedral network forming liquids. This peak includes positive contributions from all the partial structure factors. In contrast, the second peak at $q=8.2$ has positive contributions from $S_{11}$ and $S_{22}$, while the negative contribution from the partial structure factor $S_{12}$ results in a decrease of the peak height. In contrast to the strong glass former, the structure factor of the HARM system has one peak corresponding to nearest neighbor distances, Fig.~\ref{fig:Sq}b, and there is no small $q$ peak since there is no intermediate range tetrahedral order. 

For this study we examined average dynamics and dynamic heterogeneity for the strong glass former on two length scales. One length scale corresponds to motion related to the breakup of tetrahedral order, $q=5.0$, and the other length scale corresponds to motion for nearest neighbor relaxation of the particles, $q = 8.2$.
 
To be able to compare to previous work \cite{FSS}, we examine the dynamics using the overlap function,
\begin{equation}
\label{eq:O}
F_o (a;t) = \frac{1}{N} \left< \sum_n w_n(a;t) \right>,
\end{equation}
where $w_n(a;t) = \Theta[a - |\mathbf{r}_n(t) - \mathbf{r}_n (0)|]$, $\Theta$ is the Heaviside step function, and $\mathbf{r}_n (t)$ is the position of particle $n$ at time $t$. We chose two values of $a$, so that for a range of temperatures the decay time of $F_o (a;t)$ closely corresponds to the decay time of the self-intermediate scattering function, $F_s (q;t) = N^{-1} \left< \sum_n e^{-i \mathbf{q} \cdot (\mathbf{r}_n (t) - \mathbf{r}_n (0))} \right>$ at the two $q$ values of interest. We find that $F_s(q;t)$ for $q=5.0$ has approximately the same decay time as $F_o(a;t)$ for $a = 0.35$, and $F_s(q;t)$ for $q=8.2$ has approximately the same decay time as $F_o(a;t)$ for $a=0.2$. Since the plateau heights of $F_o(a;t)$ are lower than for many fragile glass formers at higher temperature, we do not use the standard definition of $\tau_\alpha$ and adapt the definitions to $F_o (a = 0.35;\tau_{\alpha}^{0.35}) = 0.2$ and $F_o (a = 0.2;\tau_{\alpha}^{0.2}) = 0.1$.

\begin{figure}
\includegraphics[scale=0.6]{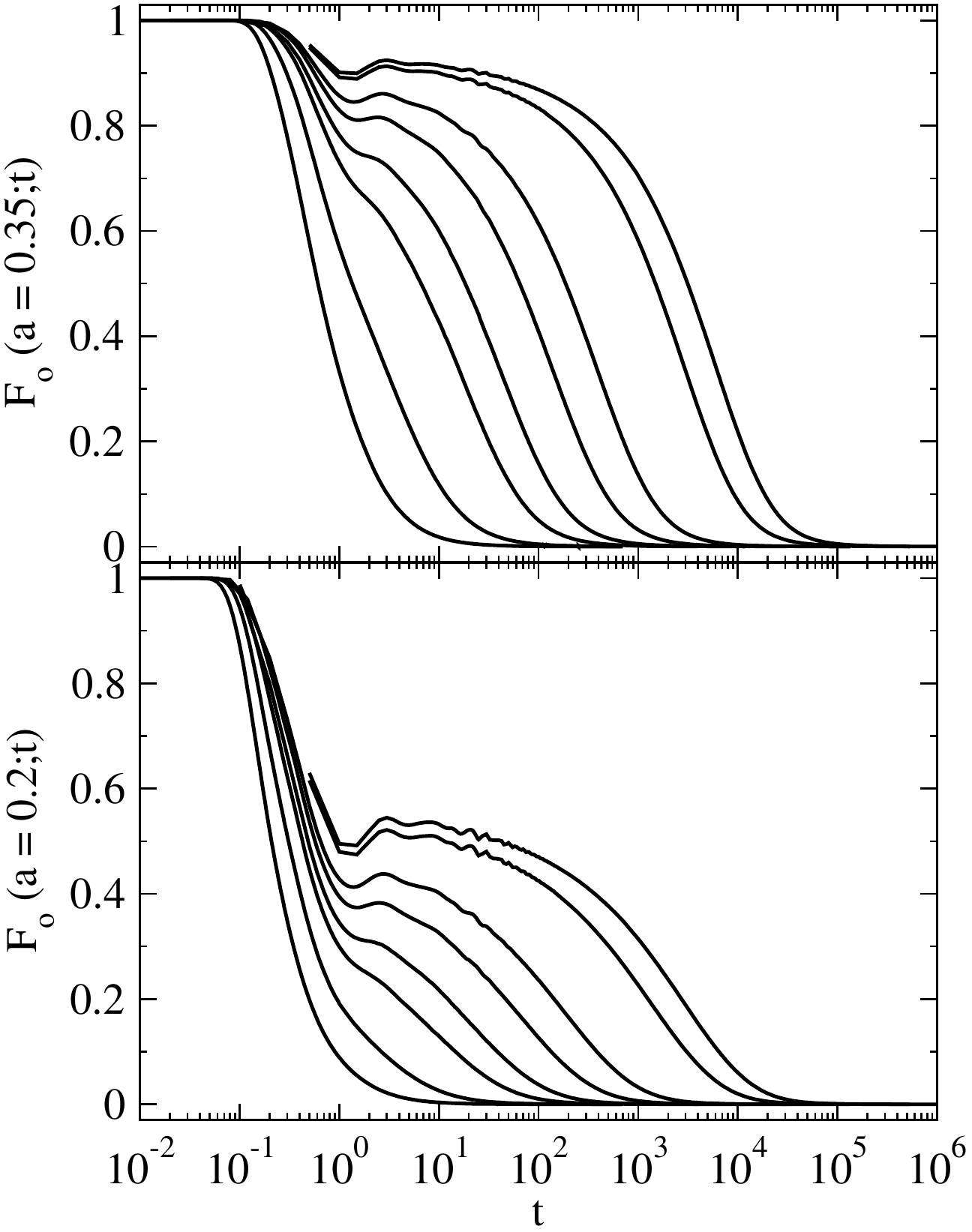}
\caption{\label{fig:theta}The average overlap function for the CP model for $a = 0.35$ (top panel) and $a = 0.2$ (bottom panel). Shown are $T=$ 0.69, 0.51, 0.42, 0.39, 0.36, 0.34, 0.31, and 0.3 listed from left to right.}
\end{figure}

Shown in Fig.~\ref{fig:theta} are average overlap functions for $a=0.35$ and $a=0.2$. For $T \le 0.42$, a plateau develops at intermediate times. The height of this plateau increases with decreasing temperature. For $1 \le t \le 100$ there are oscillations in $F_o(a;t)$ at the lowest temperatures, which we believe are due to vibrational motion.

\begin{figure}
\includegraphics[scale=0.3]{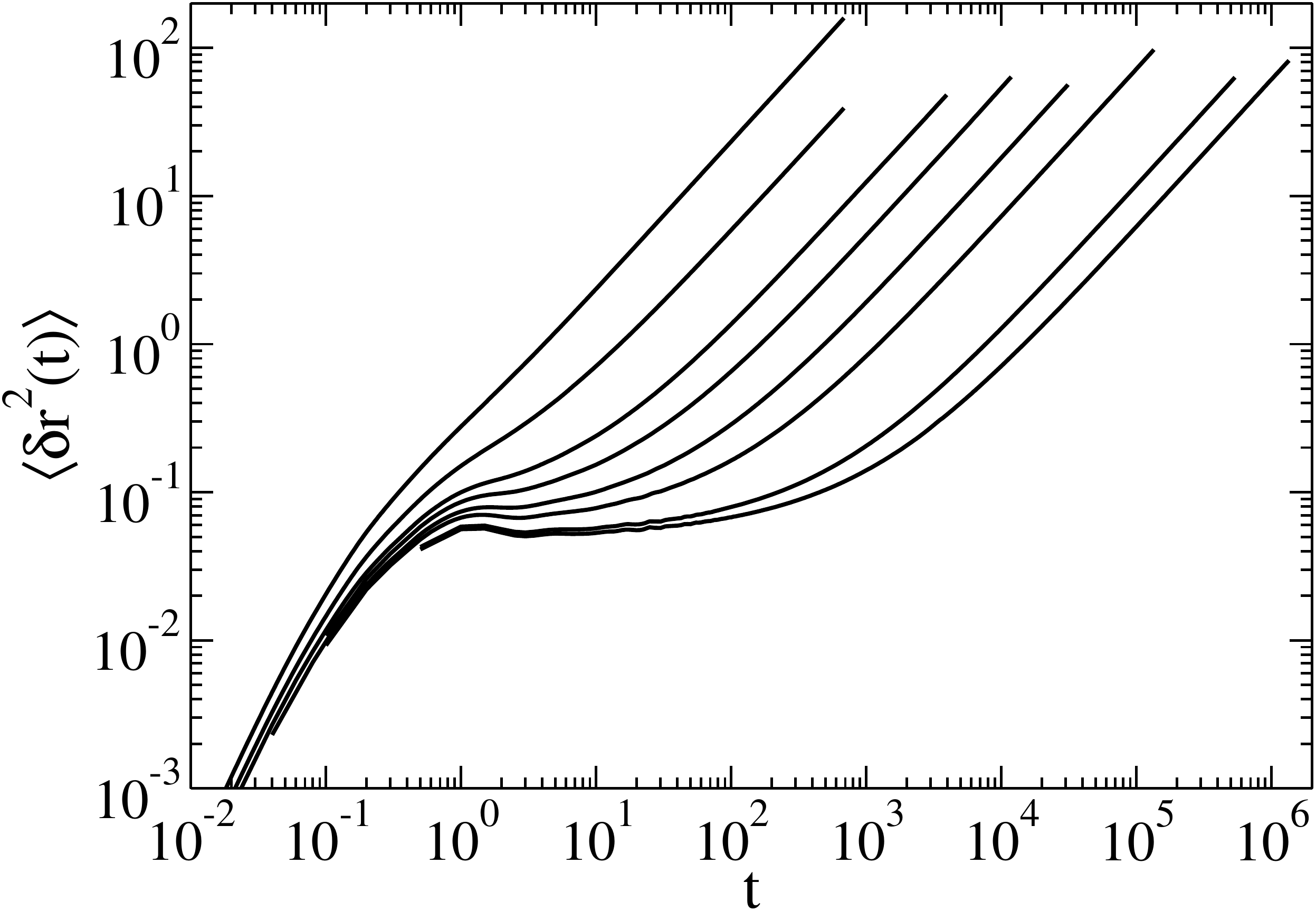}
\caption{\label{fig:msd}The mean square displacement, $\left<\delta r^2 (t)\right>$, for $T$ = 0.69, 0.51, 0.42, 0.39, 0.36, 0.34, 0.31, and 0.3 listed from left to right.}
\end{figure}

We also calculate the mean square displacement
\begin{equation}
\label{eq:msd}
\left<\delta r^2 (t)\right> = \frac{1}{N} \left< \sum_n \left[ \mathbf{r}_n (t) - \mathbf{r}_n (0) \right]^2 \right>,
\end{equation}
where $\mathbf{r}_n (t)$ is the position of particle $n$ at time $t$. Fig.~\ref{fig:msd} shows the mean square displacement for temperatures from 0.3 to 0.69. At short times the motion is ballistic, $\left<\delta r^2 (t)\right> = 3Tt^2$, and at longer times the motion is diffusive, $\left<\delta r^2(t) \right> = 6Dt$.  In the supercooled regime, $T \le 0.5$, a plateau emerges between the ballistic and diffusive regimes. We calculated the diffusion coefficient D by fitting to $\left< \delta r^2(t) \right>/(6t)$ at long times.

Finally, for temperatures higher than 0.32, we calculate the viscosity $\eta$ using, 
\begin{equation}
\label{eq:eta}
\eta = \frac{1}{k_B TV} \int_{0}^{\infty} \text{d}t \; \left< \sigma^{\alpha \beta}(t) \sigma^{\alpha \beta}(0) \right>,
\end{equation} 
where
\begin{equation}
\label{eq:sigma}
\sigma^{\alpha \beta} = \sum_{n} m_n v_n^{\alpha} v_n^{\beta} - \frac{1}{2} \sum_{n} \sum_{m \neq n} \frac{r_{nm}^{\alpha} r_{nm}^{\beta}}{r_{nm}} \frac{dU_{nm}(r_{nm})}{dr_{nm}}
\end{equation}
in the shear-stress autocorrelation function. At temperatures 0.32 and lower, our trajectories are not long enough to calculate the 
tail of the shear-stress autocorrelation function accurately. In Eq.~(\ref{eq:sigma}) $m_n$ is the mass of particle $n$, $v_n^{\alpha}$ is the $\alpha$ component of the velocity $\mathbf{v}_n$ of particle $n$, $r_{nm}^{\alpha}$ is the $\alpha$ component of $\mathbf{r}_n - \mathbf{r}_m$, and $U_{nm}$ is the potential between particles $n$ and $m$.

\begin{figure}
\includegraphics[scale=0.3]{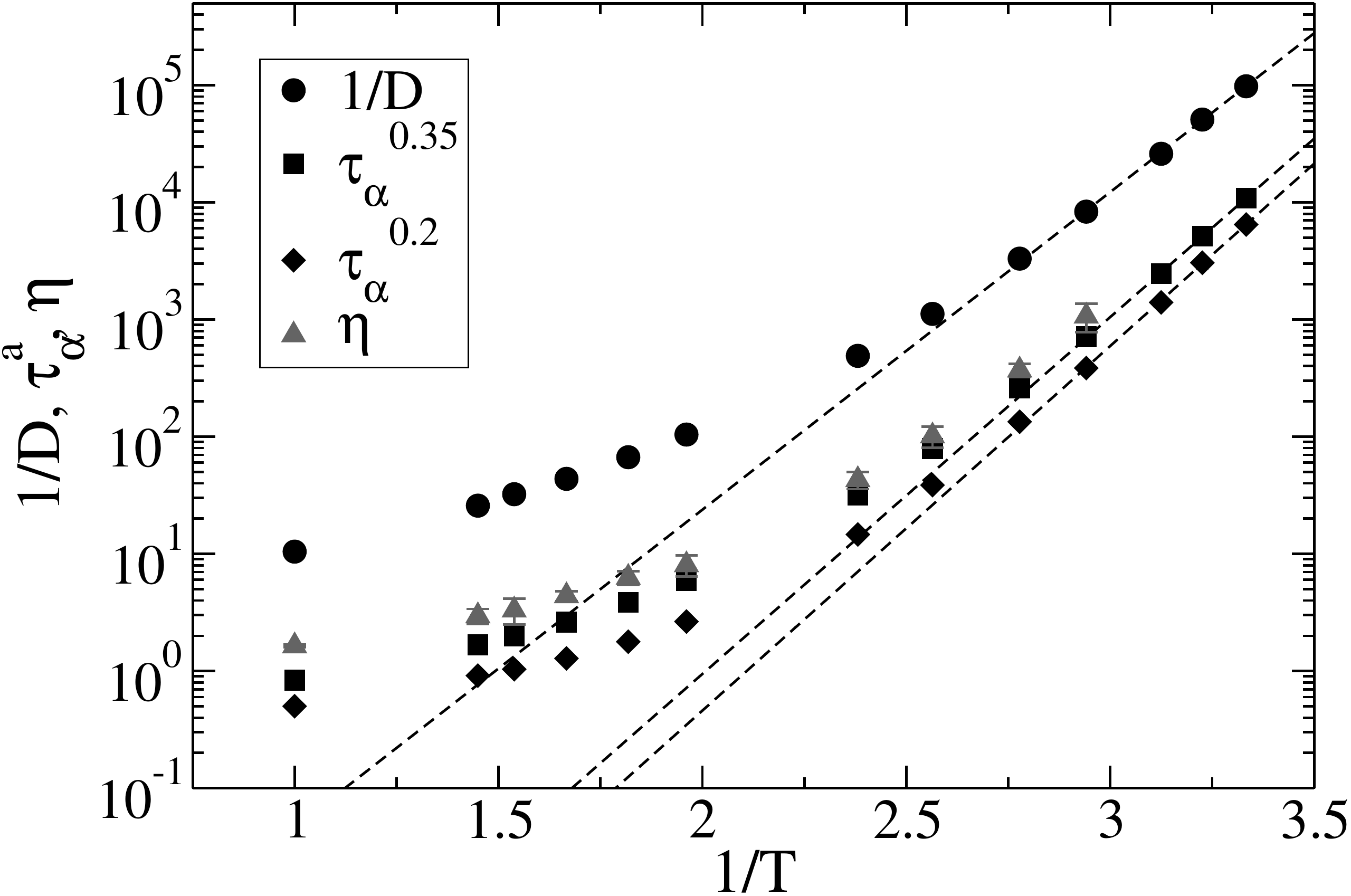}
\caption{\label{fig:Arr}The inverse diffusion coefficient $1/D$, the relaxation times $\tau_{\alpha}^a$ and the viscosity $\eta$ plotted versus inverse temperature. The black dashed lines are fits to the Arrhenius equation: $1/D = A_1 \exp{(E_1/T)}$ or $\tau_{\alpha}^a = A_2^a \exp{(E_2^a/T)}$.}
\end{figure}

Figure~\ref{fig:Arr} shows $1/D$, $\tau_\alpha^{a}$, and $\eta$ plotted versus $1/T$. The straight lines are fits of $1/D = A_1 \exp{(E_1/T)}$ and $\tau_{\alpha}^a = A_2^{a} \exp{(E_2^{a}/T)}$ to $T<0.4$. The fits were done for $T < 0.39$. The fit parameters are, $A_1 = 9 \times 10^{-5}$ and $E_1 = 6.24$. For $\tau_{\alpha}^{0.35}$, the parameters are $A_2^{0.35} = 7.7 \times 10^{-7}$ and $E_2^{0.35} = 7.01$. The parameters for $\tau_{\alpha}^{0.2}$ are $A_2^{0.2} = 2.7 \times 10^{-7}$ and $E_2^{0.2} = 7.17$. Kawasaki, Kim, and Onuki \cite{KKO}, who also simulated the CP system, found that the viscosity can be fitted to an Arrhenius equation for $T \lesssim 0.4$, and this
finding agrees with our results. Coslovich and Pastore \cite{ntw} determined a fragility index $K$
and they noted that $K = 0.09$ for the CP system, indicating that it was a stronger glass former than any previously studied Lennard-Jones mixture.

\begin{figure}
\includegraphics[scale=0.3]{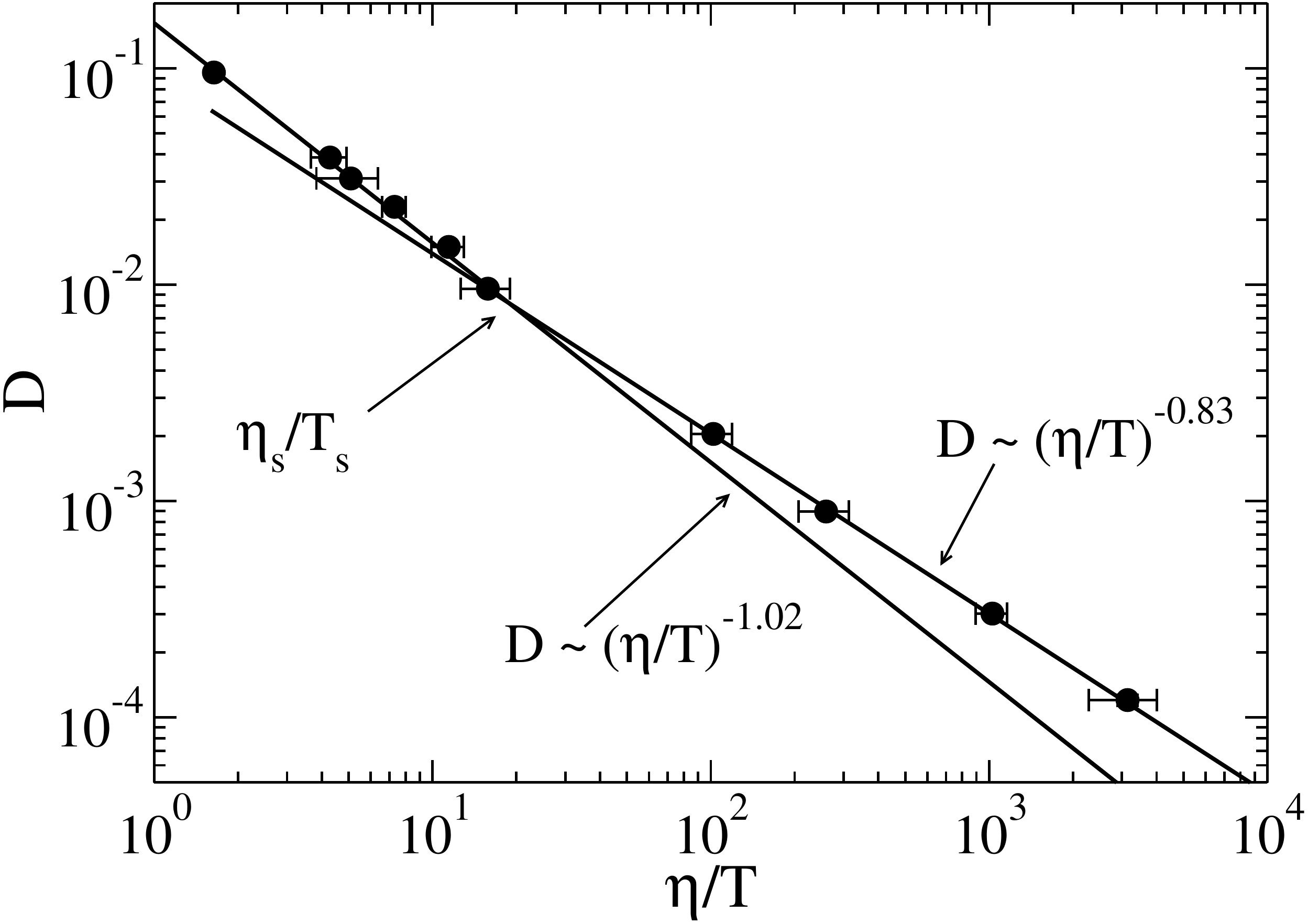}
\caption{\label{fig:D_eta}The diffusion coefficient D versus viscosity divided by temperature, $\eta/T$. Temperatures from left to right are: 1.0, 0.69, 0.65, 0.6, 0.55, 0.51, 0.42, 0.39, 0.36, and 0.34. The lines are fits to $D = c(\eta/T)^{-z}$ for $T \ge 0.51$ and for $T \le 0.42$. The fitting lines intersect at $\eta_s/T_s = 18.8$, which corresponds to a temperature $T_s = 0.5$.}
\end{figure}

To compare the strong CP system to the fragile liquids in previous work \cite{FSS}, we want to know the temperature at which the Stokes-Einstein relation $D \sim (\eta/T)^{-1}$ no longer holds. Shown in Fig.~\ref{fig:D_eta} is $D$ versus $\eta/T$. The figure also shows fits to $D = c(\eta/T)^{-z}$ for two ranges of temperatures. The fit to the higher temperatures gives $z = 1.02 \pm 0.02$ and the fit to the lower temperatures gives $z = 0.83 \pm 0.02$. The Stokes-Einstein relation is violated at $\eta_s/T_s \approx 18.8$, which corresponds to a temperature of $T_s = 0.5$. We note that the temperature where a plateau begins to emerge in $F_o(a;t)$ and $F_s(q;t)$ is approximately $T=0.5$, and the emergence of the plateau has been identified as the onset temperature for slow dynamic in Refs. \cite{ntw} and \cite{KS}.

\section{Dynamic Heterogeneity}
\label{sec:dynhet}

In previous work we found that there was a connection between Stokes-Einstein violation and the evolving shape of dynamically heterogeneous regions \cite{FSS}. Similar connections between the shapes of dynamically heterogeneous regions and Stokes-Einstein violation have been made in an experiment \cite{MG} and a simulation \cite{HBKR}. Since we found some universal behavior for at least a class of fragile glass formers, we examine whether some of these relationships hold for a model strong glass former. In the fragile glass-formers we previously studied \cite{FSS}, the dynamic susceptibility $\chi_4$, which is related to the number of particles in a region with correlated motion, is related to the dynamic correlation length $\xi_4$ by $\chi_4 \sim (\xi_4)^3$ for temperatures below the temperature at which Stokes-Einstein violation begins. This relationship suggests that regions are compact. We do not find compact regions below the Stokes-Einstein violation temperature for the CP model strong glass-former. We also find that the length scale characterizing the size of the dynamically heterogeneous regions is smaller and increases slower with decreasing temperature for the strong glass-forming liquid.

\begin{figure}
\includegraphics[scale=0.6]{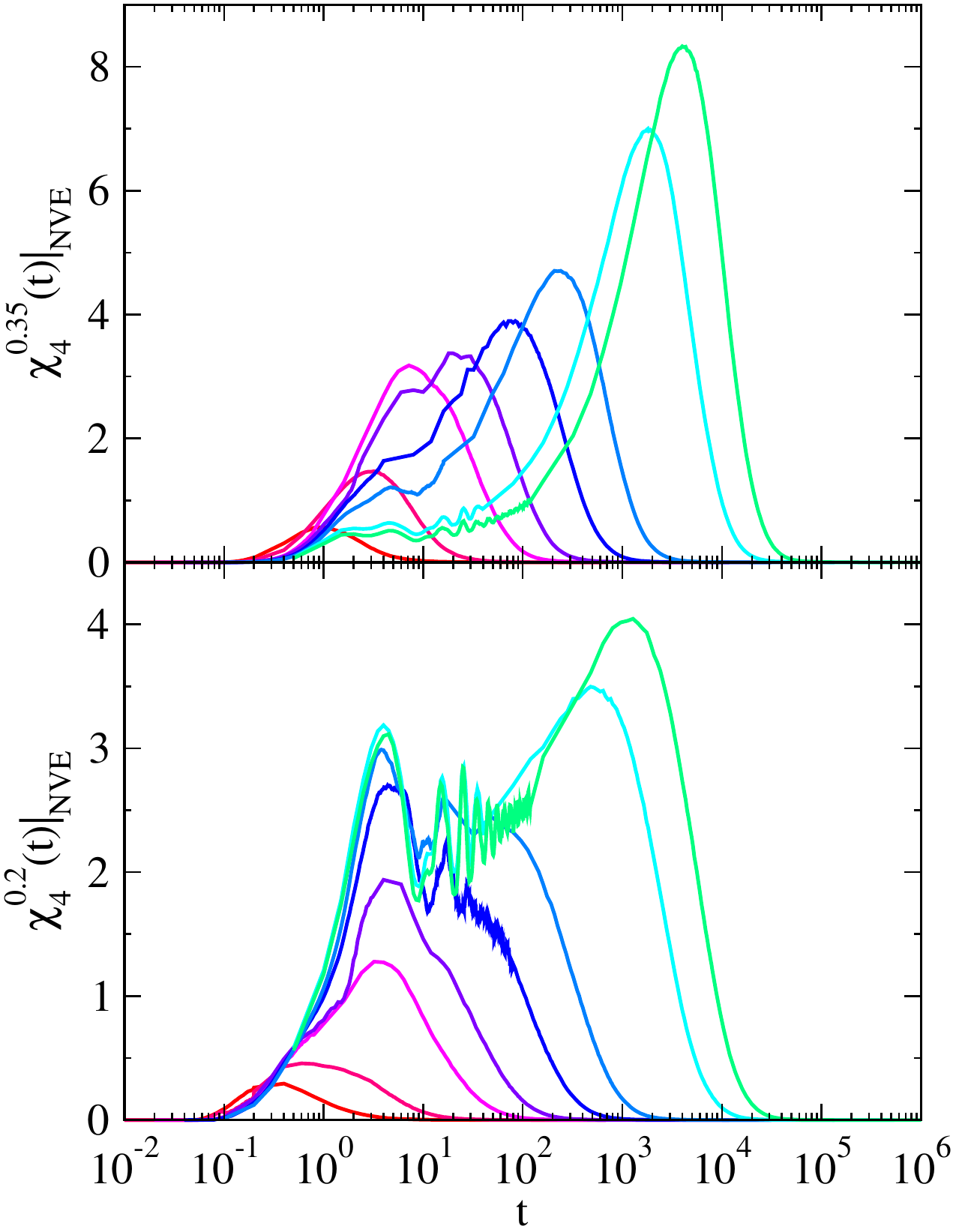}
\caption{\label{fig:X4}The susceptibility $\chi_{4}^a (t)|_{NVE}$ calculated at constant energy for $a=0.35$ (top panel) and $a=0.2$ (bottom panel). The peaks correspond to the temperatures of 0.69, 0.51, 0.42, 0.39, 0.36, 0.34, 0.31, and 0.3 listed from left to right.}
\end{figure}

First, we examine the susceptibility in the constant energy ensemble,
\begin{eqnarray}
\label{eq:X4}
\chi_4^a(t)|_{NVE} & = &\frac{1}{N} \left( \left< \left[ \sum_n w_n(a;t) \right]^2 \right> \right. 
\nonumber \\ &&
\left. - \left< \sum_n w_n(a;t) \right>^2 \right),
\end{eqnarray}
since it has features that are not present for the fragile glass formers. Shown in Fig.~\ref{fig:X4} is $\chi_4^a(t)|_{NVE}$ for the temperature range examined in this study for $a=0.35$ and $a=0.2$. $\chi_4 (t)^a|_{NVE}$ behaves differently in this network-forming liquid than in many other simulations of glass formers. For both $a$ values a side peak emerges as temperature is decreased. Similar to what was seen for $F_o(a;t)$, $\chi_4^a(t)|_{NVE}$ also has oscillations from times of around 1 to around 100. In other glass formers \cite{SF, TWBBB} the relaxation time tracks the time of the peak of $\chi_4^a(t)|_{NVE}$, $\tau_{p}^a$. In the CP system, $\tau_{\alpha}^a$ and $\tau_{p}^a$ do not have the same temperature dependence, and we found that the ratio $\tau_\alpha^a/\tau_{p}^a$ grows slightly as temperature decreases.
Due to this different behavior of $\tau_{p}^a$, we examined dynamic heterogeneity at $\tau_\alpha^a$ and at $\tau_{p}^a$.

To determine the characteristic length scale $\xi_4^a(t)$ and strength of dynamic heterogeneities $\chi_4^a(t)$ we calculated the four point structure factor
\begin{equation}
\label{eq:S4}
S_4^a(q;t) = \frac{1}{N} \left< \sum_{n,m} w_n(a;t) w_m(a;t) e^{i \mathbf{q} \cdot [\mathbf{r}_n (0) - \mathbf{r}_m (0)]} \right>.
\end{equation}
We fit $S_4^a(q;t)$ to the Ornstein-Zernicke equation $\chi_4^a(t)/[1 + (\xi_4^a(t) q)^2]$, for $q < 1.5/\xi_4^a(t)$, which is a procedure we developed in previous work \cite{FS09,FS10,FZS}. We calculated $S_4^a(q;t)$ at $\tau_{\alpha}^a$ and $\tau_{p}^a$. 

\begin{figure}
\includegraphics[scale=0.6]{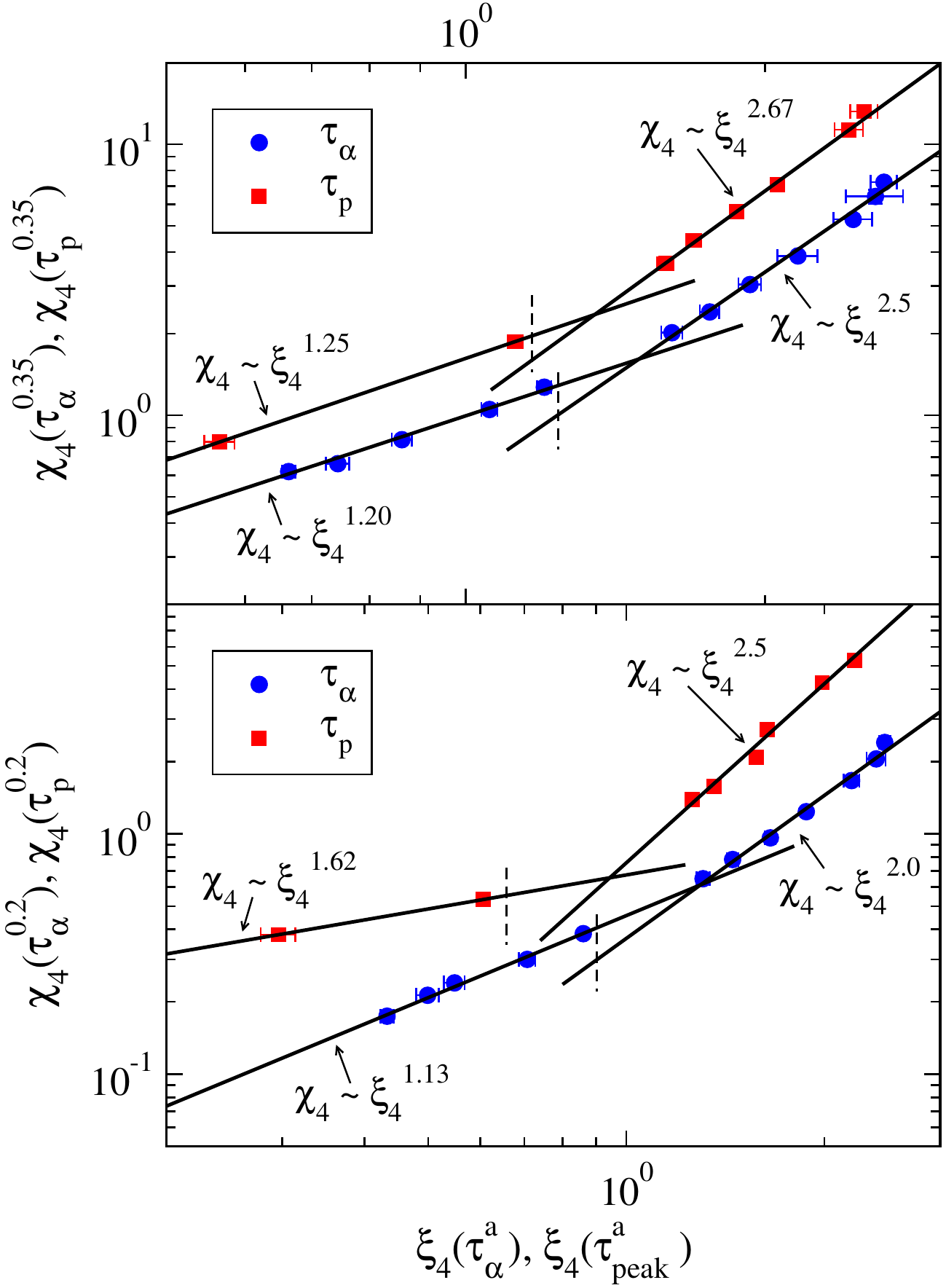}
\caption{\label{fig:X4_l_ap}The full susceptibility $\chi_4(t)$ versus dynamic correlation length $\xi_4(t)$ for two characteristic times and two characteristic length scales. The top panel shows results for $a = 0.35$, and the bottom panel shows results for $a=0.2$. The blue circles show results for $\chi_4(\tau_\alpha^a)$ and $\xi_4(\tau_\alpha^a)$, and the red squares shows results for $\chi_4(\tau_{p}^a)$ and $\xi_4(\tau_{p}^a)$. The black lines are fits to $\chi_4^a = A (\xi_4^a)^z$. The fits are to temperatures above the Stokes-Einstein violation temperature, $T_s$, and to temperatures below $T_s$. The black dashed vertical lines show the length $\xi_4^s$ that corresponds to $T_s$.}
\end{figure}

Shown in Fig.~\ref{fig:X4_l_ap} are $\chi_4^a(t)$ versus $\xi_4^a(t)$ for $t = \tau_\alpha$ and $t = \tau_{p}$ on a log-log scale. Note that there appears to be a change in power law behavior for both sets of data, and that change in behavior occurs at around the same temperature as the Stokes-Einstein violation, which is marked with a dashed vertical line. We fit $\chi_4^a = A (\xi_4^a)^z$, for values corresponding to temperatures above the Stokes-Einstein violation temperature $T_s$ and for values corresponding to temperatures below $T_s$.

As we saw for fragile liquids \cite{FSS}, Stokes-Einstein violation signals a change in the dynamic heterogeneity. For the fragile glass formers studied in Ref.~\cite{FSS}, we obtained $z = 3$ for $T < T_s$, which suggests compact clusters. The exponent in the fits for the strong glass former suggest that the clusters are more ramified than for the fragile glass formers, see Fig.~\ref{fig:X4_l_ap}.   

\begin{figure}
\includegraphics[scale=0.3]{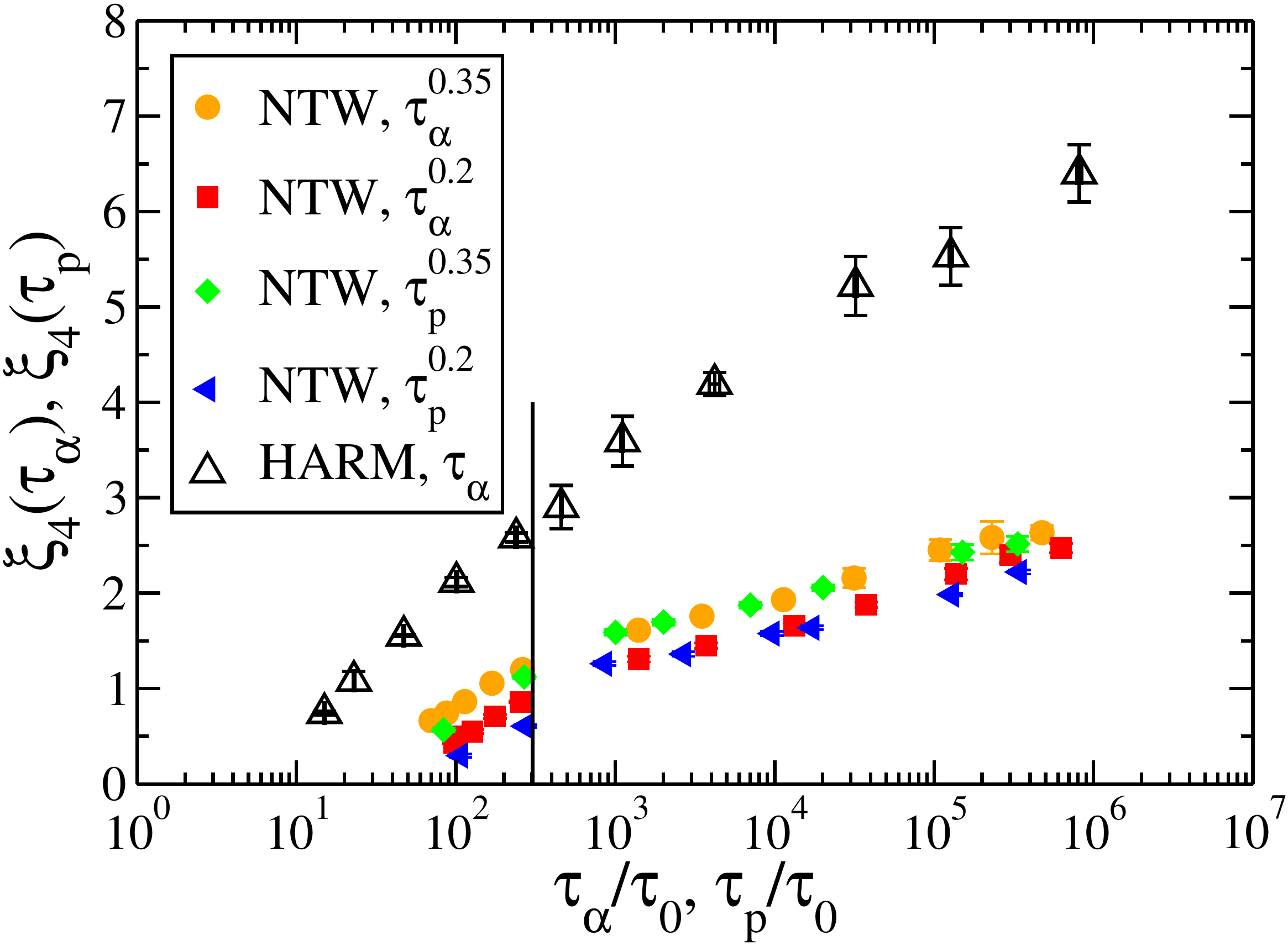}
\caption{\label{fig:l_tau}The dynamic correlation length $\xi_4$ versus rescaled time, $\tau_\alpha/\tau_0$ and $\tau_{p}/\tau_0$. The closed symbols are for the strong glass-former, the CP model, and the open symbols are for a representative fragile glass former, the harmonic sphere model. The black vertical line shows the rescaled Stokes-Einstein violation time, $\tau_\alpha^s/\tau_0$ and $\tau_p^s/\tau_0$. The temperature where Stokes-Einstein violation begins appears to mark a change in the relationship between $\tau_\alpha$ and $\xi_4(\tau_\alpha)$ and $\tau_{p}$ and $\xi_4(\tau_{p})$.}
\end{figure}

We also examined the relationship between the dynamic correlation lengths $\xi_4$ and $\tau_\alpha$ or $\tau_{p}$. Shown in Fig.~\ref{fig:l_tau} are results for the strong glass former, and we compare these results to the harmonic sphere system (triangles). In our study of fragile liquids, we found that if we rescaled the relaxation time to the relaxation time that corresponded Stokes-Einstein violation, then all the data collapsed on the same curve. We rescaled time of our strong network forming liquid so that the Stokes-Einstein violation time was 303, matching Fig. 3 of Ref.~\cite{FSS}.  

We do not find that it is possible to rescale the correlation lengths for the strong liquid and obtain reasonable collapse onto the fragile liquid data, and we also find that $\xi_4^a(t)$ for the CP model is smaller than for the fragile liquids and grows slower with decreasing temperature. (We believe that a direct comparison of the correlation lengths of the CP system and the HARM system is valid since the plateau height of the mean squared displacement had a similar value for the CP system as for the HARM system.) Since the growth is slow with decreasing temperature and the lengths are small we found it difficult to characterize the relationship between $\xi_4$ and $\tau_\alpha$ or $\tau_{p}$.

\section{Summary and Conclusions}
\label{sec:sum}

We examined dynamic heterogeneity in a model of a strong liquid. Since the four-point susceptibility, which is proportional to the number of dynamically correlated particles, can be shown to diverge through simple arguments, an interesting question is what happens to the dynamic correlation length and how this correlation length is related to the susceptibility. We calculated the dynamic susceptibility and correlation length using four-point correlation functions that are typically used to study dynamic heterogeneity in simulations. 

Since our model is a model for a network forming liquid, we studied dynamic heterogeneity corresponding to particle motion on two length scale. The shorter length scale, $a=0.2$, corresponds to relaxation at nearest neighbor distances, and the larger length scale, $a = 0.35$, corresponds to relaxation of the tetrahedral network. We find that the details change depending on the length scale studied, but the trends remain the same. There is a crossover from a high temperature behavior to low temperature behavior at the temperature where Stokes-Einstein violation begins. For the strong glass former studied here, the crossover is consistent with a change in the power law relating $\chi_4$ to $\xi_4$, i.e. a change in the exponent in the relationship $\chi_4 \sim (\xi_4)^z$. There also appears to be a change in the relationship between $\xi_4$ and $\tau_\alpha$, but the precise relationships are difficult to define due to the very slow growth of $\xi_4$. Note that this slow growth of the dynamic correlation length is likely one of the characteristics of strong glass forming systems, but one needs to study more strong glass formers to draw this conclusion. 

The change in shape of dynamically heterogeneous regions at Stokes-Einstein violation was observed in simulations of fragile glass formers \cite{FSS}. Recently, Mishra and Ganapathy \cite{MG} reported that Stokes-Einstein violation marked a change in shapes of dynamically heterogeneous regions in experiments of quasi two-dimensional ellipsoids. While two-dimensional glassy dynamics have a different character \cite{FS15NC}, it appears that Stokes-Einstein violation does mark a change in the character of dynamic heterogeneity. For the fragile systems studied in Ref.~\cite{FSS}, dynamically correlated regions were compact below the temperature where Stokes-Einstein violation begins. However, for our strong glass former, Stokes-Einstein violation does not indicate compact regions and the dynamic correlation length is still small, only around a particle diameter, at the temperature where Stokes-Einstein violation begins. 


\section{Acknowledgments}

This research utilized the CSU ISTeC Cray HPC System supported by NSF Grant CNS-0923386. We gratefully acknowledge the support of NSF grant CHE 1213401.

\end{document}